\documentclass[groupedaddress, aps, longbibliography, reprint, prl]{revtex4-1}

\usepackage{graphicx} 
\usepackage{bm}
\usepackage{bbold}
\usepackage{xcolor}
\usepackage{enumitem}
\usepackage{amsmath, amssymb, amsthm}
\usepackage[normalem]{ulem}
\usepackage{natbib}
\usepackage{comment}
\usepackage{bbm}

\newcommand{\prlsec}[1]{\textit{#1}---}

\newcommand{\ket}[1]{\ensuremath{|#1\rangle}}
\newcommand{\bra}[1]{\ensuremath{\langle #1|}}
\newcommand{\braket}[2]{\ensuremath{\langle #1|#2\rangle}}
\newcommand{\ketbra}[2]{\ensuremath{| #1\rangle\langle #2|}}

\newcommand{\sket}[1]{\ensuremath{|#1\rangle\!\rangle}}
\newcommand{\sbra}[1]{\ensuremath{\langle\!\langle #1|}}
\newcommand{\sbraket}[2]{\ensuremath{\langle\!\langle #1|#2\rangle\!\rangle}}

\newcommand{\id}{\ensuremath{\mathbb{1}}} 
\newcommand{\floor}[1]{\ensuremath{\lfloor #1 \rfloor}}

\usepackage[colorlinks=true]{hyperref}
\hypersetup{citecolor = blue, linkcolor=blue, urlcolor=blue} 

\begin{document}

\title{Provable random-matrix spectral ramp in a static, geometrically local Hamiltonian}

\author{Matteo Ippoliti}
\affiliation{Department of Physics, The University of Texas at Austin, Austin, TX 78712, USA}

\begin{abstract}
Quantum chaos is commonly associated with the emergence of random-matrix statistics in the spectra of quantum systems. A useful diagnostic is provided by the spectral form factor (SFF), which for random matrix ensembles displays a universal linear-growth regime (`ramp'). 
In the last decade, a landmark result by Bertini, Kos and Prosen (BKP)~\cite{bertini_exact_2018,bertini_random_2021} identified for the first time a class of geometrically local quantum dynamics of finite-dimensional particles where the SFF provably exhibits a random-matrix ramp: periodically driven (Floquet) qudit chains whose evolution is described by `dual-unitary' circuits. 
Here, building on the BKP result and on a recently proposed variant of the Feynman-Kitaev clock construction~\cite{ippoliti_infinite_2025}, 
we obtain a spectral ramp in a class of {\it static}, geometrically local Hamiltonians.
Our strategy is to embed the Floquet quasienergy spectrum of a dual-unitary circuit into the energy spectrum of a static local Hamiltonian, and to prove that the latter's connected SFF inherits the BKP ramp within a symmetry sector. This is to our knowledge the first proof of a spectral ramp in a time-independent, geometrically local many-body system with finite local Hilbert space dimension.
\end{abstract}

\maketitle

\prlsec{Introduction}In classical physics, chaos refers to the exponential divergence of nearby trajectories in phase space, giving rise to complex and effectively unpredictable dynamical phenomena~\cite{ott_chaos_2002}. 
An analogous notion of chaos for quantum systems is less clearly defined.
For semiclassical systems, it was conjectured that classical chaos in the $\hbar\to 0$ limit corresponds to the emergence of level repulsion described by {\it random-matrix theory} (RMT)~\cite{wigner_characteristic_1957,rosenzweig_repulsion_1960,dyson_statistical_1962,berry_level_1977,brody_random-matrix_1981} in the energy spectrum of the quantized system, whereas classically integrable systems display uncorrelated (Poisson) level statistics when quantized~\cite{casati_connection_1980,berry_quantizing_1981,bohigas_characterization_1984}.
The emergence of universal RMT statistics in the energy spectrum of many-body systems is now considered a defining feature of quantum ergodicity even in systems that lack any obvious semiclassical limit like fermions or spin-$1/2$ systems~\cite{nandkishore_many-body_2015,serbyn_spectral_2016,ho_ergodicity_2018,abanin_colloquium_2019,vikram_dynamical_2023}, 
complementing information-theoretic diagnostics based on out-of-time-ordered correlators~\cite{maldacena_bound_2016,von_keyserlingk_operator_2018,nahum_operator_2018,xu_scrambling_2024,vardhan_free_2025} or local operator entanglement~\cite{bertini_random_2026}.

A compact way to characterize RMT statistics in the energy spectrum of a Hamiltonian $H$ is through the spectral form factor (SFF)~\cite{berry_semiclassical_1985,hosur_chaos_2016,roberts_chaos_2017,kos_many-body_2018,liu_spectral_2018,cotler_chaos_2017,gharibyan_onset_2018,joshi_probing_2022}: $K_H(t) := |{\rm Tr}(e^{-iHt})|^2 = \sum_{a,b} e^{i(E_b-E_a)t}$, where $\{E_a\}$ is the spectrum of $H$ and the time $t$ sets the energy gap resolution $\sim 1/t$ being probed. Equivalently, $K_H$ is the Fourier transform of the two-point correlation function of the density of states. Averaging over Hamiltonians drawn from an RMT ensemble $\mathcal{E}$ yields the universal prediction of a linear growth in time, or `ramp': $K_{\mathcal{E}}^\mathrm{conn} (t) := \langle |{\rm Tr}(e^{-iHt})|^2\rangle_\mathcal{E}^\mathrm{conn} \propto t$, where $\langle |z|^2 \rangle^\mathrm{conn}:= \langle |z|^2\rangle - |\langle z\rangle|^2$ denotes the connected correlator. The ramp continues until the Heisenberg time, then gives way to a plateau which signals full dephasing. 

For a structured, locally interacting many-body system, the emergence of an RMT-like ramp in the SFF is a remarkable signature of universality. 
Unlike in RMT ensembles, the ramp does not set in immediately, but only after a time scale called the Thouless time, $t_\mathrm{Th}$, whose scaling with system size carries important information about the system's dynamics such as charge diffusion. 

Despite 
analytical progress in the semiclassical setting~\cite{aleiner_divergence_1997,muller_semiclassical_2004,muller_periodic-orbit_2005,pluhar_universal_2014,altland_path_2025},
in non-interacting or single-particle systems~\cite{liao_many-body_2020,winer_exponential_2020,ikeda_exact_2025}, in all-to-all-connected models such as SYK~\cite{cotler_black_2017,gharibyan_onset_2018,saad_semiclassical_2019},
and 
in the limit of large local Hilbert space dimension~\cite{chan_solution_2018,chan_spectral_2018,friedman_spectral_2019,chalker_chaotic_2025},
the only known proof of an RMT ramp in the SFF of a locally interacting system of finite-dimensional quantum particles is the result by  
Bertini, Kos and Prosen (BKP), which considered a class of periodically driven qudit chains: first the kicked Ising model at a specific self-dual point~\cite{bertini_exact_2018}, and later a broader class of driven qudit systems~\cite{bertini_random_2021} represented by `dual-unitary' quantum circuits (circuits that are unitary along both the space and time directions)~\cite{gopalakrishnan_unitary_2019,piroli_exact_2020,claeys_ergodic_2021,bertini_exactly_2026}.
For a periodically driven system, the SFF is defined in terms of the Floquet operator $U_F$, describing the unitary evolution over one period, as $K_F(w) := |{\rm Tr}(U_F^w)|^2$, with $w$ an integer counting the number of periods. BKP proved that $K_F(w) = |w|$ in the thermodynamic limit, matching exactly the RMT prediction for the circular unitary ensemble (CUE). 

Despite this achievement, an analogous example of provable RMT spectral statistics in {\it static} (as opposed to periodically driven) geometrically local Hamiltonians of finite-dimensional qudits has remained elusive. 
In this work we fill this important gap by providing such an example. Our result is based on a recent twist on the Feynman-Kitaev clock construction (a circuit-to-Hamiltonian reduction originally used in quantum complexity theory~\cite{kitaev_clock_2006}) that enables the embedding of Floquet spectra into static Hamiltonian spectra~\cite{ippoliti_infinite_2025}.
While Ref.~\cite{ippoliti_infinite_2025} focused on the implications of this embedding for eigenstates, 
here we focus on the implications for the energy level statistics. We derive a relationship between the SFF of the Hamiltonian within a specific symmetry sector and the SFF of the embedded Floquet circuit. By choosing a dual-unitary circuit, we prove that the BKP spectral ramp yields a spectral ramp in the Hamiltonian:
\begin{equation}
    \lim_{\Delta \to 0} \lim_{\tau\to\infty} \lim_{n\to\infty} 
    \frac{1}{n\tau} \overline{K_H^\mathrm{conn}}(n\tau)
    = \frac{2}{\pi},
    \label{eq:main_result_intro}
\end{equation}
where $n$ is the system size, $\tau = t/n$ is the rescaled time, $\Delta$ is the strength of disorder used to regularize the limits, and the overline denotes time averaging over a window $t\pm t\Delta$. 
The slope of $2 / \pi$ is non-universal and depends on the units of time or energy; what matters is that it is finite and non-zero.

\begin{figure}
    \centering
    \includegraphics[width=\columnwidth]{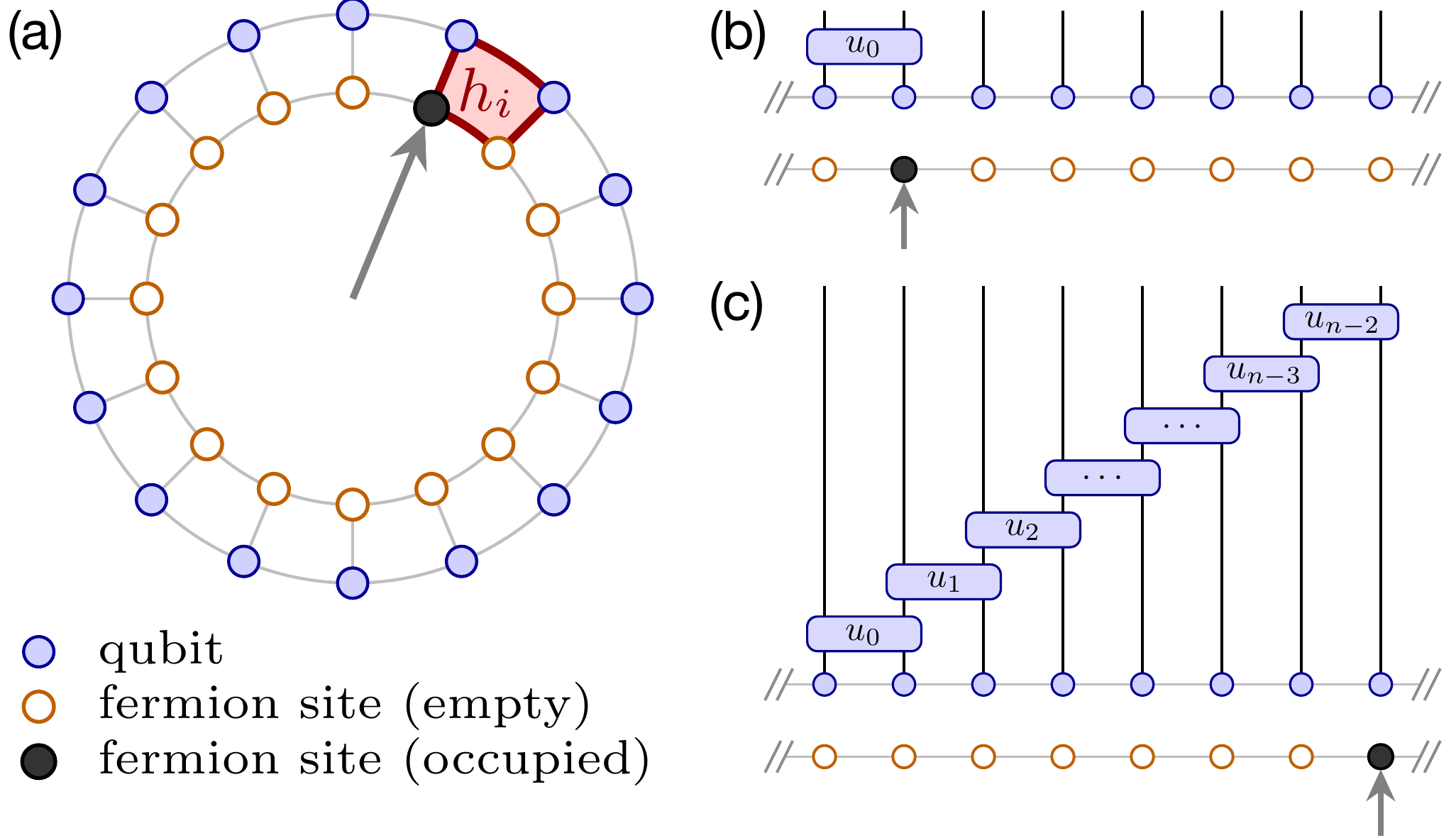}
    \caption{Schematic of the periodic Feynman-Kitaev clock Hamiltonian, Eq.~\eqref{eq:fkham}. 
    (a) The system is a two-leg ladder with periodic boundary conditions. One leg is a qubit chain, the other is a fermionic chain containing a single particle, interpreted as a clock hand (arrow). Interaction terms $h_i$ are supported on plaquettes. 
    (b) Illustration of the interaction terms: as the clock hand hops forward, it applies a gate $u_i$ to qubits $(i,i+1)$. 
    (c) Multiple hops result in a staircase circuit on the qubits. 
    In this work we always take $u_{n-1}=\id$ so the Floquet circuit $U_F:=u_{n-1}u_{n-2}\cdots u_0$ has open boundary conditions. 
    }
    \label{fig:clock}
\end{figure}

\prlsec{Review of clock construction}Here we briefly review the construction from Ref.~\cite{ippoliti_infinite_2025} which embeds Floquet quasienergy spectra into Hamiltonian energy spectra. 
Consider a two-leg ladder of length $n$ with periodic boundary conditions, Fig.~\ref{fig:clock}(a). 
One leg hosts $d$-state qudits, the other a fermionic tight-binding chain with on-site creation operators $\{c_i^\dagger\}_{i=0}^{n-1}$. In the following we stick to $d=2$ (qubits) for concreteness, though the results generalize straightforwardly to $d>2$. 
The construction takes as an input a collection of unitary gates $\{u_i\}_{i=0}^{n-1}$ on the qubit chain, with each gate $u_i$ acting on the pair of qubits $(i,i+1)$, and yields the following Hamiltonian:
\begin{equation}
    H_{FK}[\{u_i\}] = - \frac{1}{2} \sum_{i=0}^{n-1} \left( u_i \otimes c_{i+1}^\dagger c_i + u_i^\dagger \otimes c_i^\dagger c_{i+1} \right),
    \label{eq:fkham}
\end{equation}
with the convention that sites $i=0$ and $i=n$ are the same (periodic boundary conditions). 
Each interaction term $h_i := u_i \otimes c_{i+1}^\dagger c_i  + \mathrm{h.c.}$ acts on four sites (rungs $i$ and $i+1$ on the qubit-fermion ladder), making the Hamiltonian geometrically local in one spatial dimension, Fig.~\ref{fig:clock}(a). 
Throughout this work we only consider the {\it single-particle sector} of the fermion chain, $\sum_i c_i^\dagger c_i = 1$; note this is still a many-body problem overall, with Hilbert space dimension $n 2^n$. The fermion particle represents the ``hand'' of a clock whose motion activates the interactions between qubits: as the clock hops from $i$ to $i+1$, it applies gate $u_i$ to the adjacent qubits [Fig.~\ref{fig:clock}(b)]. One revolution of the hand induces one Floquet period for the qubits under the circuit $U_F := u_{n-1}\cdots u_1 u_0$ [Fig.~\ref{fig:clock}(c)]. In this work we always take $u_{n-1}=\id$ so $U_F$ has open boundary conditions. 

The Hamiltonian~\eqref{eq:fkham} was diagonalized in Ref.~\cite{ippoliti_infinite_2025}. The one-clock-hand Hilbert space decomposes into $2^n$ Krylov sectors of dimension $n$, each specified by a Floquet quasienergy $\phi$ of $U_F$ ($0\leq \phi <2\pi$). Within each sector, the Hamiltonian describes a free particle on a circle with $n$ sites threaded by an Aharonov-Bohm flux $\phi$. 
The energy spectrum is thus
\begin{equation} 
E_{m,\phi} = -\cos[(2\pi m + \phi) / n],
\label{eq:H_spec}
\end{equation}
i.e., the familiar single-particle cosine dispersion with momentum $ k = 2\pi m /n$ ($m = 0,\dots n-1$) and boundary condition twist $\phi \in \text{Spec}(U_F)$.
The quasienergy spectrum of $U_F$ is thus embedded into a real energy spectrum, where it is repeated $n$ times (one for each possible momentum of the hand) and is distorted by the cosine function.

\prlsec{Disorder averaging}Since the SFF is not self-averaging~\cite{prange_spectral_1997}, some extrinsic weak disorder must be introduced in order to get a meaningful universal answer. 
We add three sources of disorder in the problem. 
(i) We weakly deform the two-qubit gates $\{u_i\}$ entering the FK Hamiltonian, Eq.~\eqref{eq:fkham}, as $u_i \mapsto (v_{i,2} \otimes v_{i+1,1})u_i$, where each $v_{i,a}\in SU(2)$ is a weak single-qubit unitary. We parametrize them as $v_{i,a} = e^{-\frac{i}{2} \boldsymbol{\theta}_{i,a} \cdot \boldsymbol{\sigma}_i }$, in terms of three angles $\boldsymbol{\theta}_{i,a} \in\mathbb{R}^3$ for each gate ($\boldsymbol{\sigma}$ are the Pauli matrices). This choice follows BKP~\cite{bertini_random_2021}. 
(ii) We additionally apply a weak, spatially uniform phase shift $\chi$ to each of the gates $\{u_i\}$. Note that, since both $u_i$ and $u_i^\dagger$ appear in $H_{FK}$, this phase shift is physical---analogous to a Peierls phase for the clock hand. 
(iii) Finally we perform a time average: we substitute 
$t\mapsto (1+r)t$ with $r$ a small random variable to be averaged over. 

The angles $\boldsymbol{\theta}$ and $\chi$ are taken as i.i.d. normal variables of zero mean and standard deviation $\Delta>0$; 
$r$ is taken from a uniform distribution on $[-\Delta,\Delta]$. 
Thus, $\Delta$ sets the strength of all sources of disorder or averaging. It can be taken to zero only {\it after} the thermodynamic and long-time limits, as in Eq.~\eqref{eq:main_result_intro}. 
The effect of $\chi$ is to shift the phase of the Floquet unitary $U_F = u_{n-1}\cdots u_1 u_0$ as $U_F\mapsto e^{in\chi} U_F$.
Even though $\chi \sim \Delta$ is small, the overall phase shift $n\chi$ is large (recall we take $n \to\infty$ first), so $U_F$ effectively acquires a uniformly random phase factor. 
A consequence is that
\begin{equation}
    \mathbb{E}_\chi \left[ {\rm Tr}\left(U_F^{w}\right) \right]
    = e^{-\frac12 w^2 \Delta^2 n^2} \left.{\rm Tr}\left(U_F^w\right)\right|_{\chi = 0}
    \label{eq:1pt_floquet_avg}
\end{equation}
which approaches $2^n\delta_{w,0}$ at large $n$. 
An analogous result applies to the product $ {\rm Tr}\left(U_F^{w}\right)^\ast {\rm Tr}\big(U_F^{w'}\big)$, which is suppressed as $e^{-(w'- w)^2\Delta^2 n^2/2}$ after $\chi$-averaging.

\prlsec{Convolution formula for the connected SFF}We now derive one of our main results, a formula for the connected SFF of the FK Hamiltonian $H_{FK}[\{u_i\}]$ [Eq.~\eqref{eq:fkham}] in the single-clock-hand sector as a convolution of the connected SFF of the Floquet circuit $U_F = u_{n-1} \cdots u_1 u_0$. 
To ease the notation, in the following we drop the superscript from $K^\text{conn}$ with the understanding that all SFFs under discussion are connected. We also exclusively use $H$ for the restriction of Eq.~\eqref{eq:fkham} to the single-hand sector, $\sum_i c_i^\dagger c_i = 1$. 

First we define the (analytically continued) partition functions of the Hamiltonian and Floquet circuit respectively as $Z_H(it) := {\rm Tr}(e^{-iH t})$ and $Z_F(iw) := {\rm Tr}(U_F^w)$, with $t\in\mathbb R$ and $w\in\mathbb Z$. 
The partition functions are related to the connected SFFs via 
$K_H(t) = \langle |Z_H(it)|^2\rangle^\mathrm{conn}_{\boldsymbol{\theta},\chi}$ 
and 
$K_F(w) = \langle |Z_F(iw)|^2 \rangle^\mathrm{conn}_{\boldsymbol{\theta}}$.

Now we relate $Z_H$ to $Z_F$. 
Using the Jacobi-Anger identity, $e^{it\cos \theta} = \sum_{l\in\mathbb Z} i^l e^{il\theta} J_l(t)$ where the $J_l$ are Bessel functions of the first kind, and the Hamiltonian eigenvalues in Eq.~\eqref{eq:H_spec}, we write the Hamiltonian partition function as
\begin{align}
    Z_H(it)
    & = \sum_{\phi \in \mathrm{spec}(U_F)} \sum_{m=0}^{n-1} \sum_{l\in\mathbb Z} i^l J_l(t) e^{il(2\pi m + \phi)/n}.
\end{align} 
The sum over momenta $m$ yields $n$ if $l$ is a multiple of $n$, zero otherwise; 
letting $l = wn$ and recognizing the Floquet partition function, we obtain
\begin{align}
    Z_H(it)
    & = n  \sum_{w\in\mathbb Z} i^{wn} J_{wn} (t) Z_F(iw). \label{eq:zh}
\end{align} 
This expresses the Hamiltonian partition function as a linear combination of Floquet partition functions. To relate the respective SFFs, we take the absolute value squared of both sides and take the connected average over the $\boldsymbol{\theta}$ and $\chi$ disorder variables:
\begin{align}
    K_H(t)
    & = n^2 \sum_{w,w'\in\mathbb Z} i^{(w'-w)n} J_{wn} (t)^\ast J_{w'n} (t) \nonumber \\
    & \qquad \times \langle Z_F(iw)^\ast Z_F(iw') \rangle^\mathrm{conn}_{\boldsymbol{\theta},\chi} . \label{eq:kh}
\end{align} 
We first carry out the $\chi$ average, using Eq.~\eqref{eq:1pt_floquet_avg}. If $w\neq w'$, then the connected correlator is super-exponentially suppressed in $n$; the sum of those contributions vanishes in the $n\to\infty$ limit (see SI for details~\footnote{
See Supplementary Information for details on Bessel functions and the derivations of Eq.~\eqref{eq:convolution} and \eqref{eq:convolution_ramp}.
}).
This leaves the $w=w'$ terms. Recognizing $\langle |Z_F(iw)|^2\rangle^\mathrm{conn}_{\boldsymbol{\theta}}$ as the Floquet SFF, we arrive at our first main result:
\begin{align}
    K_H(t) 
    & =  2 n^2  \sum_{w = 1}^\infty  \left| J_{wn} (t)\right|^2 K_{F}(w) + o(1),
    \label{eq:convolution}
\end{align} 
where the $o(1)$ denotes terms that vanish at large system size $n$, capturing the $w\neq w'$ contributions. 
Here we used the fact that $J_{-wn}(t) = (-1)^{wn} J_{wn}(t)$ and $K_F(0) = 0$ to restrict the sum to $w>0$. 

\prlsec{Spectral ramp}To obtain a provable spectral ramp from Eq.~\eqref{eq:convolution}, we take the gates $\{u_i\}$ to be dual-unitary and not equal to the SWAP gate. 
BKP~\cite{bertini_random_2021} proved that, for a brickwork circuit of such gates on an $n$-qubit chain with periodic boundary conditions, the Floquet SFF in the thermodynamic limit is $K_F^\infty(w):= \lim_{n\to\infty} K_F(w) = |w|$ for any integer $w$~\cite{bertini_random_2021}. Before we can plug their result into Eq.~\eqref{eq:convolution}, however, we must address some differences between circuit architectures. 

Instead of a brickwork circuit, the clock construction naturally yields a staircase circuit, Fig.~\ref{fig:clock}(c).
We show in the End Matter that this is not a fundamental issue, since the staircase circuit $u_{n-1}u_{n-2}\cdots u_1 u_0$ and its brickwork counterpart $\big( \prod_{j\text{ odd}} u_j \big) \big( \prod_{j\text{ even}} u_{j}\big)$ are related by a unitary change of basis, and thus share the same spectrum and same SFF, provided the circuits have {\it open} boundary conditions ($u_{n-1} = \id$)~\footnote{
Note that even though the clock chain is always periodic (i.e., $H_{FK}$ in Eq.~\eqref{eq:fkham} contains hopping terms $c^\dagger_{n-1} c_0$ and $c^\dagger_0 c_{n-1}$ connecting the two ends of the chain), setting $u_{n-1} = \id$ gives the underlying Floquet circuit $U_F$ open boundary conditions, since no gate directly couples qubits $n-1$ and $0$.
}. 
Open boundary conditions violate the assumptions of BKP, but again this is not an issue. In the End Matter we generalize BKP's result from periodic to open boundary conditions and show that the result is unchanged up to an exponentially small correction: the limit $K_F^\infty$ exists and obeys $w\geq K_F^\infty(w) \geq w(1-w 2^{-w})$. 

With these issues addressed, we can now proceed to take the thermodynamic limit in Eq.~\eqref{eq:convolution}. 
We are interested in the scaling limit $n\to\infty$ with $t/n:=\tau$ held constant [see Eq.~\eqref{eq:main_result_intro}], also known as the Thouless double-scaling limit~\cite{chan_many-body_2022,shivam_many-body_2023}.  
A spectral ramp arises if the limit $\kappa_H(\tau) := \lim_{n\to\infty}{K_H(n\tau)} / {n}$ (which we call {\it reduced} SFF)
exists, is finite, and is linear in $\tau$. 
For the limit to exist, we must now smear the time variable. We do so as described earlier by taking $t\mapsto (1+r)t$, or equivalently $\tau\mapsto (1+r)\tau$, with $r\in[-\Delta,\Delta]$ uniformly random; $\Delta$ will be taken to zero at the end. We denote this time average by an overline. 

\begin{figure}
    \centering
    \includegraphics[width=\columnwidth]{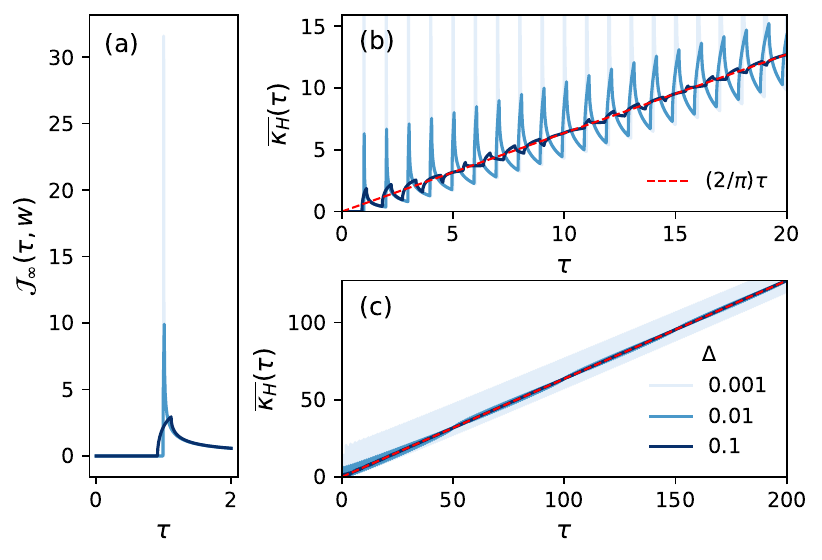}
    \caption{(a) Kernel $\mathcal{J}_\infty$ from Eq.~\eqref{eq:convolution_ramp} for $w=1$ and different time-averaging parameters $\Delta$. The peak height scales as $\Delta^{-1/2}$. (b) Time-averaged reduced SFF $\overline{\kappa_H}(\tau)$ from Eq.~\eqref{eq:convolution_ramp}, for the same values of $\Delta$. A linear ramp $\frac{2}{\pi}\tau$ (red dashed line) is shown for comparison. 
    (c) Same quantity on a longer time scale. 
    }
    \label{fig:ramp}
\end{figure}

We show in the SI~\cite{Note1} that it is safe to take the limit $n\to\infty$ inside the sum in  Eq.~\eqref{eq:convolution}. Doing so yields the reduced SFF as
\begin{align}
    \overline{\kappa_H}(\tau)
    & = 2\sum_{w=1}^\infty  K_F^\infty(w) \mathcal{J}_\infty(w,\tau),     \label{eq:convolution_ramp} 
\end{align} 
where we defined $\mathcal{J}_\infty := \lim_{n\to\infty} \overline{\mathcal{J}_n}$ and $\mathcal{J}_n(w,\tau) := n\left| J_{wn} (n\tau) \right|^2$. 
The behavior of the kernel $\mathcal J_n$ is dictated by the large-$n$ asymptotics of the Bessel function~\cite{DLMF}, which changes drastically between regions $\tau>w$ (oscillatory) and $\tau<w$ (evanescent), see SI~\cite{Note1}. Time-averaging further broadens the turning point $\tau=w$ to a crossover interval $|\tau-w| < \Delta\tau$. 
The behavior in each region is as follows.
(i) For $\tau-w>\Delta\tau$, $\mathcal{J}_n$ is given to leading order in $n$ by $\frac{2}{\pi} (\tau^2-w^2)^{-1/2} \cos^2 [n\Phi(\tau,w)]$, where the function $\Phi(\tau,w)$ is smooth with $|\partial_\tau \Phi| > \Delta^{1/2}>0$ everywhere in the region (see SI for more details~\cite{Note1}). Thus, under time averaging, the oscillatory factor $\cos^2(n\Phi)$ averages out to $1/2$ at large $n$ for any $\Delta>0$.
This gives $\mathcal{J}_\infty = \frac{1}{\pi}(\tau^2-w^2)^{-1/2}$. 
(ii) For $\tau -  w < -\Delta \tau $, $J_{nw}(n\tau)$ is exponentially suppressed in $n$, so $\mathcal{J}_\infty = 0$.  
(iii) 
Near the turning point, $|\tau-w|<\Delta \tau$,
the background singularity $\mathcal{J}_\infty \sim (\tau-w)^{-1/2}$ as $\tau\to w^+$ is cut off by the time-averaging, giving $\mathcal{J}_\infty = O(\tau^{-1}\Delta^{-1/2})$, see Fig.~\ref{fig:ramp}(a).
Note also that, exactly at the turning point $\tau=w$, $\mathcal{J}_n$ has a peak of height $O(n^{1/3})$ and width $O(n^{-2/3})$ in $\tau$; this is an integrable singularity that vanishes in the $n\to\infty$ limit after time averaging.

Combining the above contributions, the time-averaged reduced SFF is
\begin{equation}
    \overline{\kappa_H}(\tau) = \frac{2}{\pi} \sum_{w=1}^{\floor{\tau(1-\Delta)}} \frac{w}{\sqrt{\tau^2-w^2}} + O\left(\tau \Delta^{1/2}\right). \label{eq:kappa_sum}
\end{equation}
The $O(\tau\Delta^{1/2})$ error comes from the terms in the turning point region, $\floor{\tau(1-\Delta)} < w \leq \floor{\tau}$; each such term is bounded above by $O(\Delta^{-1/2})$ and there are $\Delta \tau$ such terms. 
The open-boundary correction to the Floquet SFF, i.e., replacing $w\mapsto w + O(w^2 2^{-w})$, yields an $O(1/\tau)$ correction which is subleading compared to $O(\tau\Delta^{1/2})$. 

Now we bound the scaling of the sum in Eq.~\eqref{eq:kappa_sum} with an integral. 
Noting that $f(w):= w/\sqrt{\tau^2-w^2}$ is a monotonically increasing function of $w$ in the relevant summation range, we have
\begin{equation}
    \int_0^w dw' f(w') \leq \sum_{w'=1}^{w} f(w') \leq f(w) + \int_0^{w} dw' f(w')
\end{equation}
for any integer $w>0$. 
The integral yields $\tau - \sqrt{\tau^2-w^2}$, which evaluated at $w = \floor{\tau(1-\Delta)}$ gives $\tau + O(\tau \Delta^{1/2})$; the boundary term $f(\floor{\tau(1-\Delta)})$ is $O(\Delta^{-1/2})$ which is a subleading error in the relevant order of limits $\tau \gg 1/\Delta \gg 1$. Overall we obtain
\begin{equation}
    \overline{\kappa_H}(\tau)
    = \frac{2}{\pi} \tau + O\left(\tau \Delta^{1/2} \right).
    \label{eq:sff_lim_n}
\end{equation}
We can now take the late-time limit, 
\begin{equation}
    \lim_{\tau\to\infty} \frac{\overline{\kappa_H} (\tau)}{\tau}
    = \frac{2}{\pi} + O\left( \Delta^{1/2} \right),
    \label{eq:sff_lim_ntau}
\end{equation}
and lastly the zero-disorder limit $\Delta\to 0$ to complete our proof of Eq.~\eqref{eq:main_result_intro}.

In Fig.~\ref{fig:ramp}(b,c), we numerically evaluate Eq.~\eqref{eq:convolution_ramp} at different values of the time-averaging parameter $\Delta$, confirming the predicted ramp at long times. The ramp coexists with features near integer $\tau$ associated to revolutions of the clock hand, which are smeared out under time averaging.

\prlsec{Discussion and outlook}We have proven that the periodic Feynman-Kitaev clock Hamiltonian~\eqref{eq:fkham}, when constructed from interacting dual-unitary gates and restricted to the single-clock-hand sector, exhibits a ramp in the connected SFF. This is to our knowledge the first proof of an RMT-like spectral ramp in a static, geometrically local Hamiltonian with finite-dimensional degrees of freedom. Below we offer some remarks on different aspects of the result and open questions for future work. 

First we comment on the Thouless time $t_\mathrm{Th}$ in our model. Eq.~\eqref{eq:convolution_ramp} shows that, if $\tau<1$ (i.e., $t<n$), the SFF vanishes asymptotically, giving a lower bound $t_\mathrm{Th} \geq n$. 
A linear-in-$n$ lower bound on $t_\mathrm{Th}$ is expected based on the ballistic motion of the clock hand, which needs to travel around the circumference before the system can conceivably start to exhibit global RMT signatures. Eq.~\eqref{eq:convolution_ramp} also shows that the bound is tight in our case, $t_\mathrm{Th} \propto n$, since any faster $n$-dependence would destroy the ramp in the reduced SFF $\kappa_H(\tau)$. 
For other choices of gates $\{u_i\}$, the Thouless time may be parametrically longer than $n$---e.g., by a factor of $\sim \log n$ for generic circuits~\cite{chan_spectral_2018,gharibyan_onset_2018} and by a factor of $\sim n^2$ for charge-conserving circuits~\cite{friedman_spectral_2019}. Proving this requires a different scaling limit and is left for future work. 
We note that, unlike Floquet dynamics, geometrically local Hamiltonians cannot have $t_\mathrm{Th} = O(1)$ due to energy conservation and associated hydrodynamic modes~\cite{friedman_spectral_2019,winer_hydrodynamic_2022}. The ballistic scaling found here is thus the fastest possible.

The ramp in our problem emerges in the {\it connected} SFF. The disconnected piece, $|\langle {\rm Tr}(e^{-iHt})\rangle_{\boldsymbol{\theta},\chi}|^2$, is readily calculated from Eq.~\eqref{eq:zh} and equals $4^n n^2 J_0(t)^2 \sim 4^n n^2/t$. This term decays slowly and makes the ramp invisible in the full (connected $+$ disconnected) SFF---the ramp would remain below the disconnected background even at the Heisenberg time $t\approx n2^n$. 
The slowly-decaying disconnected SFF originates from the van Hove singularities at the single-particle band edges, where the density of states diverges as $\sim |E-E_0|^{-1/2}$. 
The same edge singularity also appears in the Wishart-Laguerre RMT ensemble~\cite{liu_spectral_2018} and in supersymmetric SYK models~\cite{hunterjones_chaos_2018}, which causes the disconnected SFF to decay as $\sim D^2/t$ ($D$ being the Hilbert space dimension) and thus mask the ramp all the way to the Heisenberg time $t\sim D$. The GUE, with the spectral edge $\sim |E-E_0|^{1/2}$ from the semicircle law, gives a disconnected SFF $\sim D^2/t^3$ which masks the ramp until $t\simeq D^{1/2}$~\cite{liu_spectral_2018}. 
Taking the connected SFF reduces this sensitivity to band-edge features and yields the cleanest comparison to RMT.

An important aspect of our result is the restriction to the single-clock-hand sector, of dimension $n2^n$. While exponentially large, this sector is quadratically smaller than the full $4^n$-dimensional Hilbert space. 
Other clock hand number sectors, particularly at finite density, are beyond reach of the tools used here.
However, a nonrigorous argument (see End Matter) suggests the scaling $K_H \sim (2^n/\sqrt{n})t$ when including all number sectors: a ramp from the chaotic spin system and an exponentially large prefactor from the free-fermion dynamics of clock hands. A rigorous proof of this conjecture is a goal for future work. 
Likewise, the search for local Hamiltonians with a provable spectral ramp across the entire Hilbert space (as opposed to a symmetry sector) remains an outstanding question.

\begin{acknowledgments}
{\it Acknowledgments.} 
I thank David Long for previous collaboration on related work~\cite{ippoliti_infinite_2025} and Amos Chan and Nick Hunter-Jones for helpful comments on a draft of this manuscript. I acknowledge the use of GPT-5.5 (OpenAI) for assistance with technical derivations, literature search, and proofreading of the manuscript. 
This work was partially supported by NSF CAREER Award No. 2542880. 
\end{acknowledgments}

\bibliography{sff}

\section*{End matter}

\prlsec{From staircase to brickwork}
The SFF calculation in BKP~\cite{bertini_random_2021} applies to brickwork circuits of dual-unitary gates; the brickwork structure is essential for the space-time duality approach~\cite{bertini_exactly_2026}, where the large-size ($n\to\infty$) and finite-depth ($w = O(1)$) regime maps to a leading-eigenvalue problem for a finite-sized transfer matrix, which is tractable. 
Our embedding of Floquet circuits into static Hamiltonians, based on the periodic FK clock~\cite{ippoliti_infinite_2025}, instead applies more naturally to staircase circuits [Fig.~\ref{fig:clock}(c)]. 
Here we show that this is not an issue for the purpose of calculating the SFF, because the brickwork and staircase circuits (made from the same gates and with open boundary conditions) are related by a unitary change of basis and thus have the same spectrum. 

\begin{figure}
    \centering
    \includegraphics[width=\columnwidth]{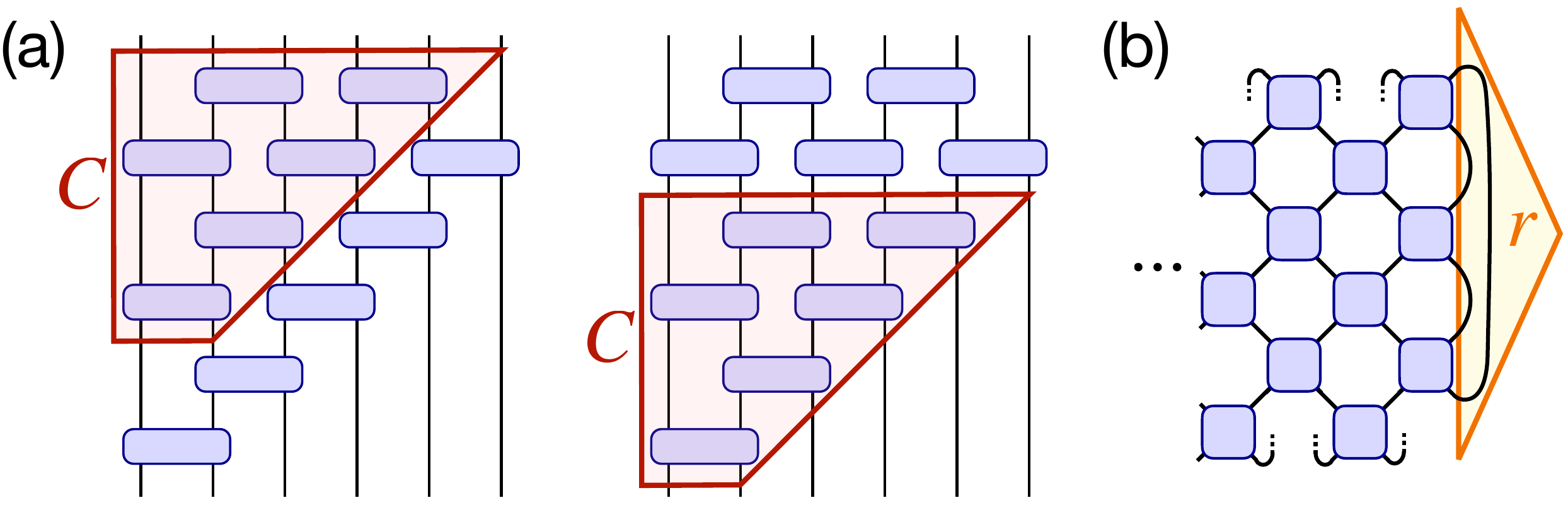}
    \caption{(a) Unitary equivalence between staircase and brickwork circuits, Eq.~\eqref{eq:brick_vs_stair}, illustrated for $n=6$ qubits. The same circuit can be expressed as $C U_{n:0}$ (left) or $U_\text{brick} C$ (right), where the circuit $C$ (shaded) is as in Eq.~\eqref{eq:change_of_basis}.
    (b) Brickwork unitary circuit with open boundary condition, shown for $w=3$ periods. The open boundary can be interpreted as an input state for the spacetime-dual dynamics, $\ket{r}$ (shaded triangular region), consisting of $w$ EPR pairs between consecutive qubits [Eq.~\eqref{eq:r_state}]. }
    \label{fig:endmatter}
\end{figure}

To show this, let us first define the partial staircase circuits $U_{b:a} := u_{b-1} \cdots u_{a+1} u_a$ for all $b>a$, so that $U_{n:0}$ is the full staircase~\cite{ippoliti_infinite_2025}. Note that we impose $u_{n-1} = \id$ to enforce open boundary conditions (no gate couples qubits $n-1$ and $0$). We also define $U_{a:a}:= \id$ and $U_{a:b} := U_{b:a}^\dagger$, so that the composition rule $U_{c:b} U_{b:a} = U_{c:a}$ holds. 
Then, the brickwork circuit 
$
U_\text{brick} := \prod_{j \text{ odd}} u_j \prod_{j \text{ even}} u_{j}
$
can be written as $U_\text{brick} = U_{2:0} U_{4:2} \cdots U_{n:n-2}$ (we assume $n$ is even). 
Now, writing $U_{4:2} = U_{4:0} U_{0:2}$, we note that $U_{0:2}$ commutes with all the unitaries to its right (this crucially requires open boundary conditions, i.e., $u_{n-1}=\id$, so that $U_{n:n-2} = u_{n-1}u_{n-2}$ acts trivially on site 0). 
Commuting $U_{0:2}$ all the way to the right, we obtain $U_\text{brick} = U_{2:0} U_{4:0} (U_{6:4}\cdots U_{n:n-2}) U_{2:0}^\dagger$. 
Iteration of this argument yields
\begin{align}
    U_\text{brick} & = C U_{n:0} C^\dagger, \label{eq:brick_vs_stair} \\ 
    C & = U_{2:0} U_{4:0} \cdots U_{n-2:0}. 
    \label{eq:change_of_basis}
\end{align}
This identity is illustrated diagrammatically in Fig.~\ref{fig:endmatter}(a). It shows that the brickwork and staircase circuits (made from the same gates $\{u_i\}$) are indeed isospectral assuming open boundary conditions. 

Note also that one could alternatively adapt the FK Hamiltonian construction to directly embed a brickwork circuit with open boundary conditions. This can be accomplished by making each gate $u_i$ act on pairs of qubits $(i+1,i+2)$ if $i$ is even, or $(i-1,i)$ if $i$ is odd, as opposed to our construction where $u_i$ always acts on $(i,i+1)$. This results in a brickwork circuit at the expense of increasing the interaction range by one.

\prlsec{Floquet SFF with open boundaries}As seen above, our embedding of brickwork circuit spectra into Hamiltonian spectra requires open boundary conditions, while BKP's results~\cite{bertini_random_2021} are obtained under periodic boundary conditions. Here we generalize their results to open boundary conditions and show that the outcome is almost identical. 

BKP show that the space-wise transfer matrix $\mathbb{T}$ of the Floquet SFF $|\mathrm{Tr}(U_F^w)|^2$ for a dual-unitary brickwork circuit $U_\text{brick}$, averaged over the $\boldsymbol{\theta}$ angles, has a $w$-fold degenerate leading eigenvalue $\lambda_0 = 1$ while the rest of the eigenvalues are strictly smaller than 1 in magnitude.
The $\lambda_0$ eigenspace is spanned by even-length translations, $\hat{T}^{2k}$, $k=0,\dots w-1$, taken with periodic boundary conditions (representing the trace ${\rm Tr}(U_\text{brick}^w)$ entering the SFF). Thus in the thermodynamic limit one obtains the projector onto even-length translations, 
\begin{equation}
    \Pi_w := \lim_{n\to\infty} \mathbb{T}^n = \sum_{a,b=0}^{w-1} \sket{\hat{T}^{2a}} (G^{-1})_{a,b} \sbra{\hat{T}^{2b}}.
\end{equation}
This is a super-operator acting on the $2w$-qubit operator space; the doubled ket/bra symbols represent vectorized operators and $G_{a,b} = \sbraket{\hat{T}^{2a}} {\hat{T}^{2b}} = {\rm Tr}(\hat{T}^{2(b-a)})$ is the matrix of overlaps between the distinct translations. 
We have $G_{a,b} = 4^{\mathrm{gcd}(w,b-a)}$, where the greatest common divisor counts the number of cycles in the translation $i \mapsto i + (b-a) \mod w$; note we define $\mathrm{gcd}(w,0) = w$.

At this point, with periodic boundary conditions as in BKP, one takes the trace: $K_F^\infty(w) = \mathrm{Tr}(\Pi_w) = \sum_{a,b=0}^{w-1} (G^{-1})_{a,b} G_{b,a} = w$ (simply the rank of the projector). 
With open boundary conditions, instead, we must use specific left and right vectors representing the configuration of tensor legs at the boundaries: 
\begin{equation}
    K^\infty_F(w) = \sum_{a,b=0}^{w-1} \sbraket{l}{\hat{T}^{2a}} (G^{-1})_{a,b} \sbraket{\hat{T}^{2b}}{r}
\end{equation}
with $\sket{r}$ the vectorization of the projector $ \ketbra{r}{r}$ and $\ket{r}$ a tensor product of (unnormalized) EPR pair states: 
\begin{equation} 
\braket{\mathbf z}{r} = \prod_{j=0}^{w-1} \delta_{z_{2j}, z_{2j\pm 1}},
\label{eq:r_state}
\end{equation}
see Fig.~\ref{fig:endmatter}(b). 
Here $\pm$ represents the two possible sublattice parities of the boundary position, resulting in different EPR pairings of the qubits. 
$\sket{l}$ is defined analogously. 
For all $a$, we have $\sbraket{l}{\hat{T}^{2a}} = \bra{l} \hat{T}^{2a} \ket{l} = \braket{l}{l} = 2^w$, since the pairing of qubits in $\ket{l}$ is invariant under the two-site translation $\hat{T}^2$; the same holds for $r$. As a result, we have
\begin{equation}
    K^\infty_F(w) = 4^w  \sum_{a,b=0}^{w-1} (G^{-1})_{a,b} 
    = w4^w \sum_{b'=0}^{w-1} (G^{-1})_{0,b'},
\end{equation}
where for the second equality we used that $G^{-1}$, like $G$, is a circulant matrix (entries depend only on the difference $b-a$). 
We can compute the sum exactly with the following trick. Introduce the uniform vector $\mathbf u$, with components $u_a := 1$ for all $a\in \{0,\dots w-1\}$. 
Noting that $\sum_b G_{a,b} u_b = \sum_{b'} 4^{\mathrm{gcd}(w,b')} := \mu_0$ is independent of $a$, we have $G\mathbf u = \mu_0 \mathbf u$. 
It follows that $G^{-1}\mathbf u = \mathbf u / \mu_0$,
which in components, and using the circulant property of $G^{-1}$, gives
$\sum_{b'} (G^{-1})_{0,b'} = 1 / \mu_0$. 
We conclude
\begin{equation}
    K^\infty_F(w) 
    = \frac{w}{\sum_{a=0}^{w-1} 4^{\mathrm{gcd}(w,a)-w}}.
\end{equation}
Now the denominator is a sum of $w$ terms. One of them ($a=0$) equals 1. 
The other $w-1$ are at most $4^{-w/2} = 2^{-w}$ (since $\mathrm{gcd}(w,a) \leq w/2$ if $a$ is not zero modulo $w$). It follows that 
\begin{equation}
w - w^2 2^{-w} \leq K^\infty_F(w) \leq w.
\end{equation}
At large $w$ the two bounds converge to the BKP result, $K_F^\infty(w) = w$.

\prlsec{Many clock hands}Here we provide a nonrigorous derivation of the Hamiltonian SFF for any number $0<M<n$ of clock hands.

We start from the energy spectrum of the $M$-hand sector, $E_{\mathbf m, \phi} = -\sum_{j=1}^M \cos[(2\pi m_j + \phi)/n]$, where $\phi\in\mathrm{spec}(U_F)$ and $\mathbf m = \{ m_1 < m_2 < \dots < m_M\}$ is a collection of occupied free-fermion momentum states~\cite{ippoliti_infinite_2025}. 
At large $n$, we approximate $E_{\mathbf m,\phi} \simeq E_{\mathbf m,0} + \rho V_{\mathbf m} \phi $, with $\rho := M/n$ the hand density, $v_m := \sin(2\pi m/n)$ the group velocity, and $V_{\mathbf m} := \frac{1}{M} \sum_{j=1}^M v_{m_j}$ the average velocity of occupied states. 
The SFF in the $M$-hand sector reads
\begin{equation}
    K_H^{(M)} (t) \simeq \sum_{\mathbf m, \mathbf m'} e^{i t(E_{\mathbf m',0} - E_{\mathbf m,0})} \sum_{\phi,\phi'} e^{it\rho (V_{\mathbf m'} \phi' - V_{\mathbf m} \phi)}.
\end{equation}
Upon averaging over $\chi$ (overall phase of $U_F$), all terms with $V_{\mathbf m'} \neq V_{\mathbf m}$ are suppressed. 
Out of the possible solutions $(\mathbf m, \mathbf m')$ to the constraint $V_{\mathbf m} = V_{\mathbf m'}$, the diagonal $\mathbf m = \mathbf m'$ receives constructive interference, while any other accidental or symmetry-based solutions dephase over time in the $e^{it(E_{\mathbf m,0} - E_{\mathbf m',0})}$ sum. Retaining only the diagonal solution leaves
\begin{equation}
    K_H^{(M)} (t) \simeq \sum_{\mathbf m} \sum_{\phi,\phi'} e^{it \rho V_{\mathbf m} (\phi'-\phi)} \simeq t\rho \sum_{\mathbf m} |V_{\mathbf m}| ,
\end{equation}
where we recognized the Floquet SFF evaluated at a non-integer argument $t \rho V_{\mathbf m} $ and substituted the ramp $K_F(w) = |w|$ (a nonrigorous step). 
This recovers a spectral ramp, although the slope is generally divergent in system size $n$. Calling $S_M$ the set of all $M$-hand configurations $\mathbf m$, the slope is $\rho \binom{n}{M} \mathbb{E}_{\mathbf m\sim S_M} |V_{\mathbf m}|$. 
This has an intuitive physical interpretation: $\rho \mathbb{E}_{\mathbf m\sim S_M} |V_{\mathbf m}|$ is the average current of clock hands, setting the conversion rate from physical time $t$ to Floquet time $w$; the binomial coefficient is the number of times each Floquet gap $\phi'-\phi$ is repeated across the Hamiltonian spectrum in the relevant number sector. 

For $M=1$, this recovers the result of Eq.~\eqref{eq:main_result_intro}: the density is $\rho = 1/n$ and the set $S_1$ is simply $\{m=0,\dots n-1\}$, giving
\begin{align}
    \frac{1}{t} K_H^{(1)}(t)
    & \simeq \frac{1}{n}\sum_{m=0}^{n-1} |\sin(2\pi m/n)| \nonumber \\
    & \simeq \frac{1}{2\pi} \int_0^{2\pi} dk\ |\sin(k)|
    = \frac{2}{\pi}. 
\end{align}
The quantitative agreement of the slope with our rigorous result, despite the various approximations, is worth noting.
For constant $M>1$, we anticipate the scaling $\sim n^{M-1} t$.
For finite density $0<\rho<1$, the average velocity of the occupied fermion states $\mathbb{E}_{\mathbf m} |V_\mathbf m|$ goes from finite to vanishing as $\sim n^{-1/2}$ (due to central-limit type sign cancellation). 
Adding over all number sectors, we expect
\begin{align}
    \frac{1}{t} K_H^\text{(all)}(t)
    & \simeq \sum_{M=1}^{n-1} \frac12 \binom{n}{M} \frac{c}{\sqrt n} = \frac{c}{2} \frac{2^n}{\sqrt n}, 
\end{align}
where, due to concentration of the binomial's weight, we replaced $\rho \mapsto 1/2$ and $\mathbb{E}_{\mathbf m}|V_{\mathbf m}| \mapsto c/\sqrt{n}$, using the values at half filling ($M=n/2$). 

\clearpage
\widetext

\setcounter{equation}{0}
\setcounter{figure}{0}
\setcounter{table}{0}
\setcounter{page}{1}
\makeatletter
\renewcommand{\thesection}{S\arabic{section}}
\renewcommand{\theequation}{S\arabic{equation}}
\renewcommand{\thefigure}{S\arabic{figure}}

\section{Supplementary information}

\subsection{Bessel function facts} 

Here we recapitulate some facts about the Bessel functions of the first kind, $J_\nu(t)$, that are used in this work~\cite{DLMF}. 

\begin{itemize}
    \item \uline{Definition}. $J_\nu(t)$ is given by the series expansion
    \begin{equation}
        J_\nu(t) = \sum_{k=0}^\infty (-1)^k \frac{(t/2)^{2k+\nu}}{k!\Gamma(\nu+k+1)}
    \end{equation}
    with $\Gamma$ the Euler gamma function. 
    \item \uline{Symmetry.}
    $J_{-\nu}(t) = J_{\nu}(-t) = (-1)^\nu J_{\nu}(t)$.
    \item \uline{Uniform upper bound.} 
    For all integer $\nu\geq 1$ and all $t\in \mathbb R$, the absolute value $|J_\nu(t)|$ obeys 
    \begin{equation} 
    |J_\nu(t)| \leq \frac{1}{\nu!} (|t|/2)^\nu \leq \left(\frac{e |t|}{2\nu} \right)^\nu 
    \label{eq:uniform_bound}
    \end{equation} 
    \item \uline{Large-$n$ asymptotics}. At fixed $\tau = t/n$, $w = \nu / n$, and large $n$, we have the following scalings up to subleading-in-$n$ corrections.
    \begin{itemize}
    \item Forbidden region $w>\tau$:
    \begin{equation}
        |J_{nw}(n\tau)|^2 \sim \frac{\exp[-2nw(\alpha-\tanh\alpha)]}{2\pi nw\tanh \alpha},
        \qquad
        \alpha = \mathrm{arccosh}(w/\tau)>0.
    \end{equation}
    In particular, $|J_{nw}(n\tau)|^2$ decays exponentially in $n$ at fixed $w>\tau$ (note $\alpha -\tanh\alpha > 0$). 
    \item Turning point $w=\tau$: in a window $|\tau-w|=O(n^{-2/3})$, 
    $J_{nw}(n\tau)$ is described by Airy scaling.
    For $\tau = w + s w^{1/3}n^{-2/3}$, with $s=O(1)$,
    \begin{equation}
        |J_{nw}(nw+s(nw)^{1/3})|^2
        =
        \left(\frac{2}{nw}\right)^{2/3}
        |\mathrm{Ai}(-2^{1/3}s)|^2
        +
        O(n^{-1})
    \end{equation}
    with $\mathrm{Ai}$ the Airy function. 
    In particular, the turning-point peak has height $O(n^{-2/3})$ and width $O(n^{-2/3})$ in $\tau$.
    \item Allowed region $w<\tau$:
    \begin{equation}
        |J_{nw}(n\tau)|^2 \sim \frac{2}{\pi n \sqrt{\tau^2-w^2}} \cos^2[n\Phi(\tau,w) - \pi/4],
        \qquad
        \Phi(\tau,w) =  \sqrt{\tau^2-w^2} - w \arccos(w/\tau)
    \end{equation}
    The function $\Phi$ is smooth everywhere in $\tau > w$ and its derivative $\partial_\tau \Phi = \sqrt{1-(w/\tau)^2}$ does not vanish. In particular, if $\tau-w> \Delta \tau$, then $w/\tau < 1-\Delta$ and $|\partial_\tau\Phi| > \sqrt{2\Delta-\Delta^2} > \Delta^{1/2}$ at small $\Delta$.  
    \end{itemize}
\end{itemize}

\subsection{Bounding $w\neq w'$ terms in Eq.~\eqref{eq:convolution}}
Here we show that all terms with $w\neq w'$ in Eq.~\eqref{eq:kh} add up to an $o(1)$ value, as reported in Eq.~\eqref{eq:convolution}. 
First, we show the following bound:
\begin{align} 
|\langle Z_F(iw)^* Z_F(iw')\rangle^\text{conn}_{\boldsymbol{\theta},\chi}| 
& \leq e^{-\frac{1}{2} \Delta^2 (w-w')^2 n^2} |\langle Z_F(iw)^* Z_F(iw')\rangle^\text{conn}_{\boldsymbol{\theta}}| \nonumber \\
& \leq e^{-\frac{1}{2} \Delta^2 (w-w')^2 n^2} 
\left( |\langle Z_F(iw)^* Z_F(iw')\rangle_{\boldsymbol{\theta}}|
+ |\langle Z_F(iw)\rangle_{\boldsymbol{\theta}}^* \langle  Z_F(iw')\rangle_{\boldsymbol{\theta}} |\right) \nonumber \\
& \leq e^{-\frac{1}{2} \Delta^2 (w-w')^2 n^2} 
\left( \langle |Z_F(iw)| |Z_F(iw')| \rangle_{\boldsymbol{\theta}}
+ \langle |Z_F(iw)| \rangle_{\boldsymbol{\theta}} \langle |Z_F(iw')| \rangle_{\boldsymbol{\theta}}\right)  \nonumber \\
& \leq e^{-\frac{1}{2} \Delta^2 (w-w')^2 n^2} 2^{2n+1}\,. 
\label{eq:bound_zf_zf}
\end{align}
Here we used Eq.~\eqref{eq:1pt_floquet_avg} for the $\chi$ average and then the fact that $|Z_F| \leq 2^n$. 
With this, we can bound the sum of $w\neq w'$ terms using the triangle inequality as follows:
\begin{align}
\left| \sum_{w\neq w'} J_{nw}(t)^* J_{nw'}(t) \langle Z_F(iw)^* Z_F(iw')\rangle^\text{conn}_{\boldsymbol{\theta},\chi} \right|
& \leq \sum_{w\neq w'} |J_{nw}(t)^*J_{nw'}(t)| 2^{2n+1} e^{-\frac{1}{2} \Delta^2 (w-w')^2 n^2} \nonumber \\
& \leq 2^{2n+1} e^{-\frac{1}{2} \Delta^2 n^2} \sum_{w, w'} |J_{nw}(t)^*J_{nw'}(t)| \nonumber \\
& \leq 2^{2n+1} e^{-\frac{1}{2} \Delta^2 n^2} \left( |J_0(t)| + 2\sum_{w=1}^\infty \frac{1}{(nw)!} \left(\frac{|t|}{2}\right)^{nw} \right)^2 \nonumber \\
& \leq 2^{2n+3} e^{-\frac{1}{2} \Delta^2 n^2} e^{|t|}
\end{align}
In the first line we used the bound on the connected correlator in Eq.~\eqref{eq:bound_zf_zf}; 
in the second line we used $(w-w')^2\geq 1$ and then reintroduced the $w=w'$ terms (note the argument of the sum is non-negative); 
in the third line we used the uniform bound on $|J_{nw}(t)|$, Eq.~\eqref{eq:uniform_bound};
in the fourth line we used $|J_0(t)|\leq 2$ and bounded the sum over $w$ from above by relaxing the $\nu = nw$ constraint and recovering the Taylor series of the exponential. 
We see that, for all $\Delta>0$, the absolute value of the sum converges to zero in the $n\to\infty$ limit, as claimed.

\subsection{Exchanging limit and sum in Eq.~\eqref{eq:convolution_ramp}}
In deriving Eq.~\eqref{eq:convolution_ramp} from Eq.~\eqref{eq:convolution}, we exchange the order of the thermodynamic limit $n\to\infty$ and of the infinite sum over $w$. 
To justify this step, we split the infinite sum into a finite sum $w < W$ and a tail, $w\geq W$, with $W$ large compared to $\tau$ but independent of $n$. 
We have 
\begin{align}
    \left| \sum_{w=W}^\infty \overline{\mathcal{J}_n}(\tau,w) K_F(w) \right| 
    & \leq  \sum_{w=W}^\infty \left|  \overline{\mathcal{J}_n}(\tau,w)\right| | K_F(w) |  \nonumber \\
    & \leq \sum_{w=W}^\infty n \left(\frac{e\tau}{2w}\right)^{2nw} 4^n \nonumber \\
    & \leq 2 n \left(\frac{e\tau}{W} \right)^{2nW}.
\end{align}
In the second line we used Eq.~\eqref{eq:uniform_bound} to bound $\overline{\mathcal{J}_n}$ and also used the uniform finite-size bound $K_F(w) \leq 4^n$ for the Floquet SFF. 
In the third line we set $W$ large enough that $q:= (e\tau/W)^2 < 1/2$ and bounded the sum above by $\sum_{w=W}^\infty q^{nw} = q^{nW}/(1-q) \leq 2q^{nW}$. 
The overall upper bound is $o(1)$ in $n$, showing that the tail sum vanishes as $n\to\infty$. Therefore we have
\begin{align}
    \lim_{n\to\infty} \sum_{w=1}^\infty \overline{\mathcal{J}_n}(\tau,w) K_F(w) 
    & = \lim_{n\to\infty}  \sum_{w=1}^{W-1} \overline{\mathcal{J}_n}(\tau,w) K_F(w) 
    + \lim_{n\to\infty}  \sum_{w=W}^\infty \overline{\mathcal{J}_n}(\tau,w) K_F(w) \nonumber \\
    & = \sum_{w=1}^{W-1} \lim_{n\to\infty}  \overline{\mathcal{J}_n}(\tau,w) K_F(w) + 0 \nonumber \\
    & = \sum_{w=1}^{\infty}   \mathcal{J}_\infty (\tau,w) K_F^\infty (w),
\end{align}
where in the last line we extended the sum to $w=\infty$ again by noting that for $w\geq W>\tau$, each term $\mathcal{J}_\infty(\tau,w)K_F^\infty(w)$ equals zero.

\end{document}